# Improving municipal responsiveness through AI-powered image analysis in E-Government

**Catalin Vrabie**
*National University of Political Studies and Public Administration,
Bucharest, Romania*

*Abstract.* Integration of Machine Learning (ML) techniques into public administration marks a new and transformative era for e-government systems. While traditionally e-government studies were focusing on text-based interactions, this one explores the innovative application of ML for image analysis, an approach that enables governments to address citizen petitions more efficiently. By using image classification and object detection algorithms, the model proposed in this article supports public institutions in identifying and fast responding to evidence submitted by citizens in picture format, such as infrastructure issues, environmental concerns or other urban issues that citizens might face. The research also highlights the Jevons Paradox as a critical factor, wherein increased efficiency from the citizen side (especially using mobile platforms and apps) may generate higher demand which should lead to scalable and robust solutions. Using as a case study a Romanian municipality who provided datasets of citizen-submitted images, the author analysed and proved that ML can improve accuracy and responsiveness of public institutions. The findings suggest that adopting ML for e-petition systems can not only enhance citizen participation but also speeding up administrative processes, paving the way for more transparent and effective governance. This study contributes to the discourse on e-government 3.0 by showing the potential of Artificial Intelligence (AI) to transform public service delivery, ensuring sustainable (and scalable) solutions for the growing demands of modern urban governance.

## Introduction

The adoption of digital technologies has transformed public service delivery, evolving through stages from basic electronic systems to e-government 3.0, which integrates emerging technologies such as Artificial Intelligence, blockchain, and the Internet of Things (IoT) (Vrabie and Dumitrascu 2018). This evolution underscores the need for innovative solutions to enhance citizen engagement, service delivery, and governance.

Artificial Intelligence (AI) and Machine Learning (ML) are transformative tools for public administration, moving beyond textual data applications like chatbots to address the growing demand for image-based submissions in public service systems. This shift addresses urban challenges where visual data is increasingly used to report issues such as infrastructure damage, illegal activities, and environmental concerns (Digi24 2015, C. Vrabie 2024).

One of the central challenges faced by public administrations is balancing efficiency with the increasing volume of citizen interactions (Glaukande and Mzini 2019). The Jevons Paradox provides an insightful framework to understand this dynamic. This paradox suggests that improvements in efficiency, often enabled by technological advancements, can lead to an increase in overall demand, potentially overwhelming the very systems designed to improve performance (Alcott 2005). In the context of e-government, the ease of submitting petitions or complaints through digital platforms can result in an increased number of submissions, necessitating scalable and robust solutions to handle the growing workload. Machine Learning however, offers a promising approach to address this paradox by streamlining processes and enhancing the capacity of public administration systems.

The progression to e-government 3.0 signifies a new era where technologies like ML are not simply supportive tools but integral components of governance capable of assisting public servants in their duties. By incorporating such techniques, governments can process large volumes of data, identify patterns, and predict outcomes with greater accuracy. For instance, ML models trained on citizen-submitted images can automatically detect and classify issues such as potholes, unauthorized constructions, or public infrastructure damages. These systems not only reduce the pressure on human operators but also enable faster response times, thereby improving citizen satisfaction and trust in government institutions.

In this study, the author focuses on the application of ML for image analysis in e-government systems. Using datasets from a Romanian municipality, the author explored how ML models can enhance the efficiency and accuracy of public service delivery. The research builds on the premise that modern citizens demand prompt, effective, and transparent governance, which is often constrained by limited public budgets and administrative inefficiencies. By leveraging ML for image analysis, public administrations can overcome these limitations, ensuring that citizen concerns are addressed in a timely and effective manner.

The integration of ML into public administration also addresses the increasing complexity of citizen interactions. Unlike textual data, which requires natural language processing (NLP) for analysis, image data demands sophisticated computer vision techniques, including object detection, segmentation, and classification. These techniques enable systems to "understand" visual data in much the same way that humans do, identifying key features and categorizing them based on predefined criteria (C. Vrabie 2022). For example, a citizen might submit an image of a damaged road, which the ML model can analyze to determine the severity of the issue, its location, and the necessary response. This capability not only enhances the precision of issue identification but also facilitates the prioritization of resources for addressing critical concerns.

As the study will demonstrate, the shift to image-based e-petition systems align with the broader goals of e-government 3.0, which aims to create more inclusive, transparent, and efficient governance structures. By integrating ML into these systems, governments can achieve several key objectives such as (1) enhancing citizen participation – mobile platforms make it easier for citizens to report issues, fostering greater engagement and participation in governance (C. Vrabie 2015), (2) improving service delivery – ML models enable faster and more accurate processing of citizen submissions (Rahmat, Vrabie and Soesilo 2023), ensuring that services are delivered in a timely manner (Tirziu and Vrabie 2017) and (3) increasing transparency and trust – automated systems provide consistent and impartial responses, reducing the potential for bias or error and enhancing public trust in government institutions (Zankova 2021, Tirziu and Vrabie 2016).

To illustrate these concepts, the study utilizes a case study from a Romanian municipality, which has been at the forefront of adopting smart city technologies. The municipality provided anonymized datasets comprising citizen-submitted texts and images related to various urban issues. These datasets were processed using state-of-the-art ML models, demonstrating the potential of algorithms to enhance public service delivery. The results reveal that ML can significantly reduce the time and effort required for issue identification and resolution, while also improving the accuracy and consistency of responses.

The study contributes to the growing body of literature on e-government 3.0 and the role of emerging technologies in governance. By also focusing on image analysis, the research highlights a critical area of innovation that has received relatively little attention in the existing discourse. While much of the most recent literature on e-government emphasizes textual data and natural language processing, this study underscores the importance of expanding the scope of AI applications to include visual data. This shift is particularly relevant in the context of smart cities, where visual information often forms the basis of citizen interactions with public administration systems.

**Literature review**

The rapid evolution of e-government has been characterized by the adoption of emerging technologies that enhance public service delivery and citizen engagement. Initially, e-government systems focused on text-based interactions and basic electronic service delivery (e-government 1.0). Later, social media and Web 2.0 technologies brought a participatory dimension to governance (e-government 2.0) (Garuckas and Kaziliūnas 2008). Today, e-government 3.0 leverages cutting-edge technologies such as Artificial Intelligence (AI), Machine Learning (ML), and the Internet of Things (IoT), promising transformative improvements in governance processes (C. Vrabie 2023).

AI applications in public administration have drawn significant attention in recent literature, with many studies emphasizing their role in automating repetitive tasks and improving governance efficiency. For example, (Chen, Ahn and Wang 2023) and (Noordt and Misuraca 2022) highlight the potential of AI to enhance transparency, citizen participation, and decision-making in public administration, while (Kolkman 2020) discusses the applicability of algorithmic models for policy-making. However, much of this research has focused on text analysis, neglecting the increasing role of image analysis in governance.

Recent advancements in computer vision, a branch of AI that enables machines to interpret visual data, offer immense potential for e-government applications. Technologies such as object detection, image segmentation, and visual classification have been widely adopted in various domains, including urban planning, environmental monitoring, and public safety. For instance, (Chui, Roberts and Yee 2022) and (Ahn and Chen 2022) discuss the societal benefits of applying deep learning for social good, including its use in public infrastructure management. Similarly, (Zhao, et al. 2022) highlight the impact of digital advancements, including AI-driven visual analytics, on sustainable development goals.

The role of computer vision has been explored in specific contexts. (Kumari, Agarwal and Mittal 2021) developed cross-domain sentiment analysis models to improve human interactions, while (Yu, et al. 2022) proposed knowledge-enhancing methods for classification tasks, demonstrating their relevance to both textual and visual data. In the context of e-government, these advancements can be applied to analyze citizen-submitted images of urban issues, such as damaged infrastructure or environmental hazards, enabling automated classification and prioritization of tasks. Building upon advancements in computer vision, researchers have explored its potential in diverse applications, including urban safety diagnostics. Back in 2013, (Zaki, et al. 2013) implemented computer vision techniques to address pedestrian safety issues at a major intersection in Vancouver, Canada. Their approach involved automated analysis of video sequences to identify pedestrian-vehicle conflicts and traffic violations, such as jaywalking. This methodology enabled proactive safety evaluations, providing valuable data for developing countermeasures and improving urban planning. Similarly, deep learning models have been leveraged for slum mapping (Wurm, et al. 2019), urban green space assessments (Helbich, et al. 2019), and safety analytics, demonstrating the versatility and impact of these technologies in urban contexts

E-petitioning, a critical application in e-government, has traditionally relied on natural language processing for analyzing textual petitions. Authors such as (C. Vrabie 2023) and (Hashem, et al. 2023) have explored the role of AI in triaging and responding to petitions, emphasizing the benefits of automation in reducing administrative workload. However, the integration of image-based petitions remains an underexplored area. With the growing prevalence of mobile technologies, citizens increasingly submit images alongside or instead of textual descriptions to report issues. This trend underscores the need for integrating computer vision techniques into e-government systems to enhance their capabilities.

Several studies have demonstrated the effectiveness of ML in automating administrative processes. (Davenport and Ronanki 2018) discuss the application of AI for automating routine decision-making tasks in the public sector. Similarly, (Scholl and Bolívar 2019) emphasizes the regulatory and competitive benefits of AI adoption in governance. While these studies provide a foundation for understanding the potential of AI, they do not adequately address the specific challenges associated with visual data in governance.

Recent work in computer vision has focused on improving the accuracy and interpretability of image analysis models. For instance, Residual Neural Network (ResNet) (Gu, Bai and Kong 2022), You Only Look Once (YOLO) (Redmon, et al. 2015) and more recent Joint Embedding Predictive Architecture (JEPA) (LeCun 2024) architectures have become standard for tasks such as object detection and segmentation, enabling machines to identify patterns and anomalies in visual data with high precision. In the context of e-government, these models can be adapted to analyze citizen-submitted images, identifying issues such as potholes, unauthorized constructions, or waste disposal problems. The ability to automate these tasks can significantly enhance the efficiency of public service delivery.

Beyond technical advancements, the societal implications of AI and computer vision in governance have been a focus of recent studies. (Eom and Lee 2022) explore how digital government transformation, driven by AI, can address complex societal challenges, while (Sorrell 2009) as far as 2009 revisits the Jevons Paradox in the context of technological advancements, highlighting the potential for increased demand to offset efficiency gains. This paradox is particularly relevant to e-government systems, where the ease of submitting image-based petitions may lead to an increased number of submissions, necessitating scalable solutions.

In the context of smart cities, (Vlahovic and Vracic 2014) and (Twizeyimana and Andersson 2019) emphasize the role of digital governance in enhancing urban management. Computer vision techniques have been applied to monitor urban environments, detect anomalies, and predict maintenance needs, demonstrating their relevance to e-government applications. For instance, (Madan and Ashok 2023) discuss governance maturity as a critical factor in AI adoption, while the afore mentioned authors (Ahn and Chen 2022) highlight public employees' positive perception of AI technologies, particularly in tasks involving visual data.

The Romanian discourse on e-government 3.0, as reflected in the works published on platforms such as scrd.eu, emphasizes the transformative role of Artificial Intelligence in public administration, highlighting its potential to enhance governance efficiency and societal well-being. Scholars like (C. Vrabie 2023) underscore AI's capacity to drive innovation in smart cities, improving citizen engagement and administrative responsiveness. Meanwhile, (Sarcea 2024) delve into the critical intersection of AI and cybersecurity, advocating for robust AI-driven mechanisms to protect public sector digital infrastructures and ensure data integrity. Addressing the cultural dimensions of e-government adoption, (Iqbal and Genie 2022) alongside with (Schachtner 2021 and Schachtner 2022) highlight how societal norms and local cultural factors shape citizens' behavioral intentions toward e-government platforms, emphasizing the need for culturally sensitive and inclusive designs. Beyond governance, the discourse also explores AI's cross-sectoral impact, as seen in (Iancu, Vrabie and Ungureanu 2021), (Stankevičiūtė and Kumpikaitė-Valiūnienė 2023) and (Zota and Clim 2023), who discusses digital and AI's application in education and also healthcare to improve service delivery, illustrating its broader public good potential. This body of research reflects a forward-looking vision for Romania, where e-government 3.0, powered by AI, not only enhances administrative capabilities but also aligns with the principles of sustainability, transparency, and citizen-centric governance (Stoicescu, Bițoiu and Vrabie 2023).

Other international journals, such as, Public Policy and Administration has published several articles that delve into the integration of digital tools within public administration, highlighting its transformative potential and associated challenges. For instance, in the article "Evolution of E-government in the States of the Eastern Border of the European Union," the author analyzes the differences in e-government development among EU member states, emphasizing the role of technological advancements in enhancing public service delivery (Žilinskas 2014). Similarly, (Žilionienė 2004) provides an early overview of key documents guiding e-government development, underscoring the importance of strategic planning in integrating digital technologies into public administration. These studies underscore the overall significance of electronic tools in facilitating democratic processes and improving public service delivery. Additionally, the journal has explored the ethical and practical implications of technology deployment in governance, emphasizing the need for a balanced approach that leverages technological advancements while safeguarding public trust and ensuring accountability (Čeičytė and Petraitė 2014). These contributions provide valuable insights into the later evolving discourse on AI's role in public administration, offering guidance for policymakers and practitioners aiming to harness AI for public good.

While much of the existing research has focused on textual data, the growing prevalence of image-based citizen interactions calls for a broader perspective. This study seeks to bridge this gap by exploring the potential of computer vision techniques in e-government applications. By integrating visual analytics into e-petition systems, public administrations can enhance their capacity to process and respond to citizen concerns, ensuring more efficient and inclusive governance. The following sections build on this foundation, presenting a case study and methodology for implementing ML-driven image analysis in public administration systems.

**Materials and methods**

This study aims to demonstrate the application of Machine Learning techniques for processing and analyzing images to enhance e-government systems, particularly in handling citizen petitions. Starting from a rather old initiative, namely Civic Alert (Digi24 2015, HotNews 2016), the author is proposing an enhanced system that adapts and expands by integrating advanced computer vision and natural language processing techniques. The steps undertaken in the process are detailed as follows:

- *Data collection and anonymization* - for this study the datasets were obtained from a municipality in Romania known for its advanced smart city initiatives. The data covered a full year (2022):
  - Images (5,712 images paired with location data) - citizens submitted photographs documenting public infrastructure issues such as damaged roads, illegal parking, and waste disposal.
  - Textual data (12,935 textual entries comprising complaints, suggestions, and inquiries) - accompanying descriptions of the images and standalone textual petitions submitted via multiple channels, including emails, mobile apps, and web platforms. While this study focuses exclusively on image analysis, details regarding the text analysis methodology and findings can be found in a previously published article, which the author recommend consulting for additional insights: "E-Government 3.0: An AI Model to Use for Enhanced Local Democracies" (C. Vrabie 2023).
- *Dataset description and distribution* - the dataset was categorized into three primary classes based on the issues reported:
  - Infrastructure damage (45% - including potholes, damaged sidewalks, and broken streetlights);
  - Waste disposal (30% - improper garbage disposal, illegal dumping sites, and overflowing bins);
  - Illegal parking and miscellaneous (25% - vehicles obstructing public spaces, abandoned cars, and other urban concerns such as graffiti and fallen trees or branches).
- *Environmental and visual conditions* - the dataset was categorized into three primary classes based on the issues reported:
  - Lighting Conditions - approximately 35% of the images were taken under low-light or nighttime conditions, requiring the model to handle variations in brightness and contrast. The rest of the dataset included images captured during daylight, with varying degrees of shadows.
  - Weather conditions - around 20% of the images were taken under adverse weather conditions, including rain, snow, or fog. These factors introduced additional challenges such as reflections, blurring, and obstructions.

- o Clutter and occlusion - nearly 25% of the dataset featured cluttered scenes with multiple overlapping objects or partially visible issues, simulating real-world complexities in urban environments.
- o Seasonal variation – as mentioned, the dataset covered a full year (2022) therefore all seasons, ensuring a diverse representation of conditions such as wet roads during autumn or snow-covered infrastructure in winter.

The diversity of the dataset presented several challenges, including distinguishing between similar issues (e.g., wet spots versus potholes) and identifying issues in visually complex or low-quality images. However, these challenges mirror real-world scenarios that municipal systems encounter, making the dataset an ideal benchmark for testing the proposed solution.

- *Preprocessing* - data preprocessing ensured the datasets were clean, standardized (by TensorFlow/Keras (TensorFlow 2024)), and anonymized:
    - o Resizing, cropping and normalization - images were resized and cropped to 256x256 pixels to standardize input dimensions for ML models and normalized to fit within a [0, 1] range.
    - o De-noising - filters such as Gaussian blur were applied to reduce noise.
    - o Augmentation - techniques like rotation, flipping, and zoom were employed to increase dataset variability, improving model robustness.

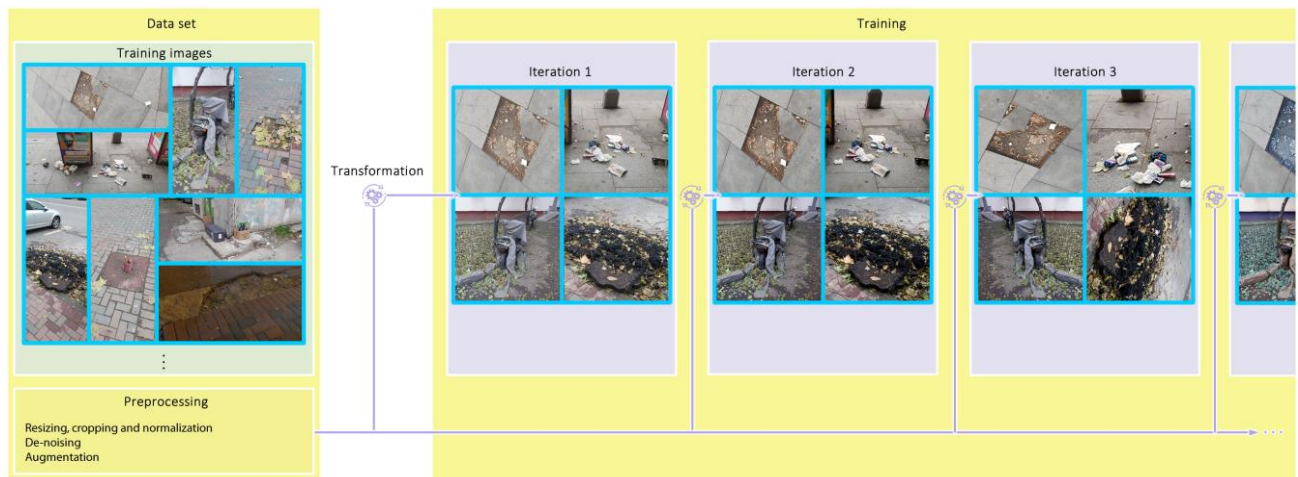

*Figure 1. **Image transformation – preprocessing (visual representation[1]**)*
*Source: Author.*

- *Feature extraction* - to enable ML algorithms to understand the input data, feature extraction techniques were applied (in this study, both Faster R-CNN and ResNet were used in complementary roles within the image analysis pipeline):
    - o Region proposals - using models like Region-based Convolutional Neural Network (Faster R-CNN), the most relevant regions of interest (ROIs) in the images were identified for issue classification. These regions were localized areas in the images that potentially contained urban issues (e.g., potholes, waste). This object detection step was crucial for analyzing complex scenes containing multiple issues or cluttered backgrounds.
    - o Convolutional features - a Convolutional Neural Network (CNN) pre-trained on ImageNet (for this study the author was using ResNet-50 (Viso.ai 2023)) was used to extract visual features such as edges, shapes, and textures. At this step the ROIs

---

[1] Computers interpret pictures as data. Images ar seen as arrays of numbers, where each pixel is represented by its coordinates (two numbers) and its color values (three numbers, one for each RGB channel). For example, a 256x256 pixel image could be represented by an array of 327,680 numbers. When computers compare images, they actually compare the arrays (transformed into vectors) corresponding to each image. In the illustration, the author provides a visual representation of the numerical data associated with each picture.

identified by Faster R-CNN were processed to extract high-level features and classify the detected issues into predefined categories such as "infrastructure damage," "illegal parking," and "waste disposal." The ResNet was fine-tuned on the dataset, as described in the methodology, for increased accuracy in identifying urban issues.

Integrating both Faster R-NN for object detection and ResNet for classification ensured a robust pipeline capable of handling both tasks in an effective manner.

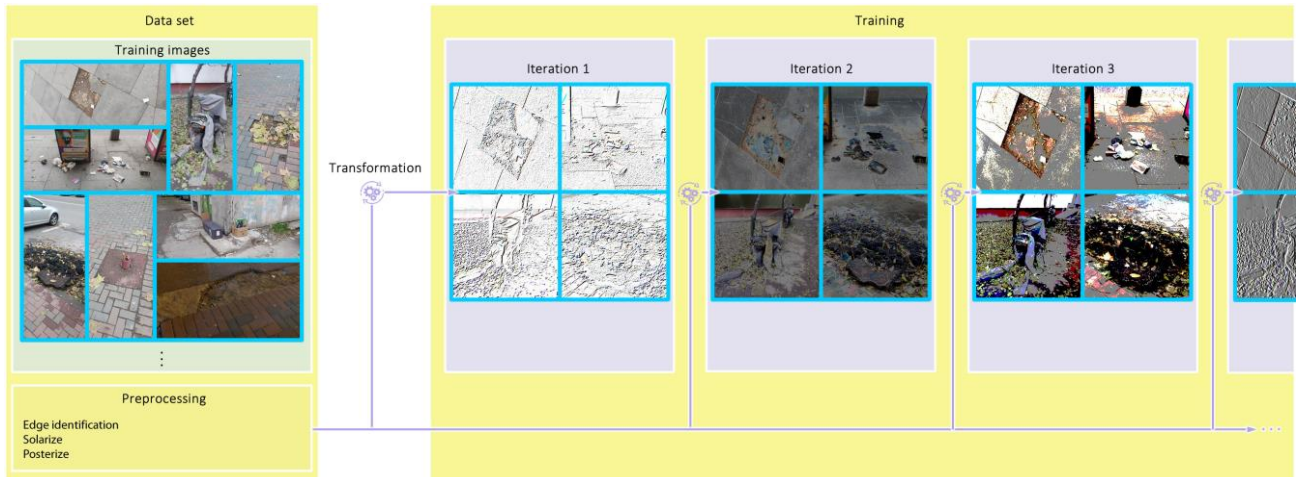

*Figure 2. **Feature extraction (visual representation)***
*Source: Author.*

- *Data annotation* - a team of experts (municipal public services analysts) manually annotated a subset of the dataset to create training and validation labels. For images, categories included "infrastructure damage," "illegal parking," and "waste disposal."
- *Model development* - a fine-tuned ResNet model was trained to classify issues based on visual features extracted from citizen-submitted images after they were preprocessed by TensorFlow/Keras. The model architecture included convolutional layers for feature extraction, followed by fully connected layers for classification.
- *Model training and validation*
    - Training - data was split into training (4570 pictures - 80%) and validation (1142 pictures - 20%) sets. Augmented datasets were used during training to prevent overfitting.
    - Hyperparameter optimization - parameters such as learning rate, batch size, and number of epochs were optimized using grid search.
    - Evaluation metrics - models were evaluated based on accuracy, precision and recall. Confusion matrices were generated to analyze misclassifications.

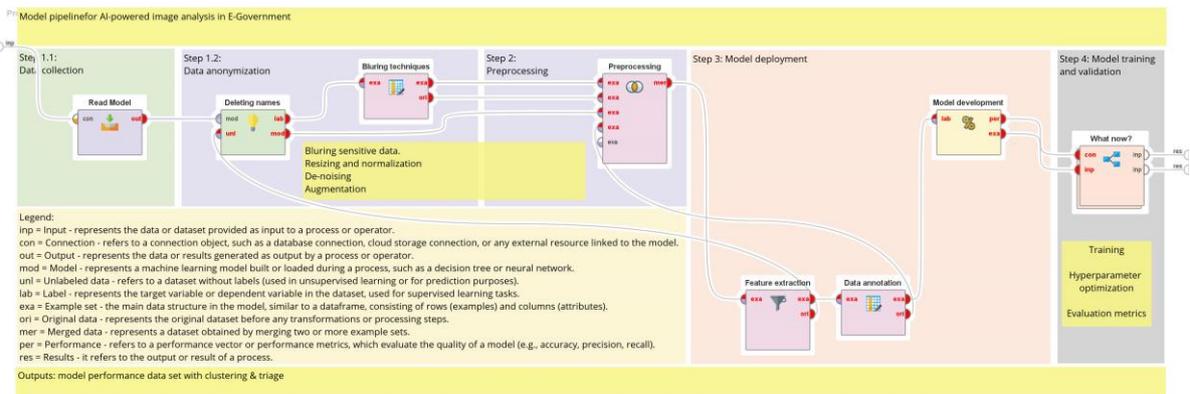

*Figure 3. **Model pipeline (visual representation)***
*Source: Author.*

**Results**

This study tested the efficiency and accuracy of an image classification model integrated into an e-government system for analyzing citizen-submitted visual data and automating subsequent actions. The proposed system processes urban issue reports by analyzing images, classifying the identified problems, generating relevant documents for responsible authorities, and promptly communicating resolutions to citizens. The results below include specific benchmarks achieved using nothing more than a Dell XPS 9510 laptop with the following specifications: 11th Generation of Intel Core i9-11900H @ 2.50 GHz CPU, equipped with an NVIDIA GeForce RTX 3050 Ti GPU, 32 GB of RAM with a running 64-bit Windows 11 Pro operating system, demonstrating its capability as the sole hardware for this application.

The image classification model, based on the Faster R-CNN architecture, was trained and evaluated using a dataset of 5,712 images submitted by citizens. The system, implemented and tested exclusively on the afore mentioned laptop, demonstrated the following capabilities:

- *Accurately identify urban issues*
    - The model effectively detected and classified common urban problems, including potholes, waste disposal issues, illegal parking, fallen branches or trees, abandoned animals, water leaks, abandoned vehicles, graffiti and acts of vandalism as well as other types of damaged public infrastructure, all grouped for this study into three primary classes: infrastructure damage, waste disposal, and illegal parking/miscellaneous issues. Metrics such as accuracy, precision, recall, and F1-score were computed for each class to evaluate performance in a more granular manner.
    - Achieved a classification accuracy (number of correct prediction / total number of predictions) of 92.8% on the validation dataset, with precision (true positives / (true positives + false positives)) and recall (true positives / (true positives + false negatives)) values of 0.93 and 0.91, respectively.
    - Misclassification rates were minimal, with errors occurring primarily in visually ambiguous images (e.g., poor lighting or highly cluttered scenes).

| CLASS | PRECISION (%) | RECALL (%) | F1-SCORE (%) |
|---|---|---|---|
| Infrastructure Damage | 94 | 92 | 93 |
| Waste Disposal | 91 | 90 | 91 |
| Illegal Parking / Miscellaneous | 89 | 85 | 87 |

These results indicate that the model performs slightly worse on the "Illegal Parking / Miscellaneous" class, likely due to higher intra-class variability and visual ambiguity in the dataset.

- *Rapid data processing*
    - On the Dell XPS 9510 laptop, each image was processed in an average of 3.1 seconds, including identifying regions of interest and generating output.
    - The complete workflow, including classification, report generation, and citizen communication, was executed in under 7 seconds per case.
- *Automated report generation* - the system integrates classification results into an automated workflow, with the following functionalities:
    - Generating actionable reports - upon detecting an issue, the model generates a report tailored to the responsible authority. These reports are formatted to comply with administrative requirements ("compare and comply" function[2]), minimizing the need for human intervention.

---

[2] This capability is able to read the legislation and, comparing with the identified issue, to refer to specific regulations in use (IBM Watson - https://www.ibm.com/docs/en/cloud-private/3.2.x?topic=paks-watson-compare-comply-element-classification).

- o Citizen answer - simultaneously, the system drafts a message for the citizen, acknowledging receipt of the report and providing details of the actions initiated using "compare and comply" function.
- *Better time management* - by automating the analysis and response processes and utilizing nothing more than the afore mentioned laptop for computation, the system demonstrated significant improvements in time management and human labor optimization. While traditional methods required an average of 8 minutes per case to analyze images, identify issues, and prepare responses, using the proposed ML system, this was reduced to less than 7 seconds per case, representing a time savings of approximately 98.5%. This efficiency would allow municipal staff to focus on strategic tasks rather than repetitive administrative duties.
- *Performance on limited hardware* - the Dell XPS 9510 laptop, handled all computations smoothly without additional external hardware. The system processed up to 500 cases per hour under normal operating conditions, demonstrating the scalability and efficiency of the setup for medium-sized municipalities.
- *Scalability* - while suitable for smaller municipalities or localized e-government services, the model supports scalable workflows, capable of handling larger data volumes with modest adjustments to batch processing or model optimization.

The results indicate that a not expensive hardware solution should be enough for implementing an ML-powered e-government system focused on image classification for small municipalities, however, as the volume increases better hardware solution should be taken into consideration. The system's ability to process reports efficiently, generate actionable outputs, and communicate with citizens demonstrates its potential to revolutionize public administration workflows. The reduction in processing time, enhanced responsiveness, and scalability makes this approach particularly suitable for municipalities aiming to enhance service quality while optimizing resource use.

**Discussion**

This study demonstrated the effective integration of machine learning techniques, particularly image classification, into e-government workflows for handling urban issues. By leveraging a robust system architecture powered by widely available hardware, such as the Dell XPS 9510 laptop, the proposed solution achieved significant advancements in time efficiency and labor optimization predicting an increased citizen engagement. The system's ability to process images, generate actionable reports, and communications with citizens highlights its potential to enhance public service delivery in a scalable and cost-effective manner.

Machine Learning in e-government plays a pivotal role by automating administrative practices, streamlining workflows, and enhancing citizen trust through precise and timely feedback. These improvements to the initial project – Civic Alert, demonstrate the transformative role that AI-powered systems can play in modern governance.

The system's scalability demonstrates AI's capacity to support data-driven governance and optimize resource allocation. By efficiently processing large data volumes and integrating metadata analysis, it provides actionable insights essential for urban planning and addressing high service demand.

However, implementing such advanced technologies requires addressing practical and ethical challenges, including hardware capabilities, system optimization, citizen readiness, and avoiding disparities in technology access. The study reflects AI's broader applicability to public administration theory, offering a framework for scalable implementation in governance and contributing to the discourse on data-driven and ethical decision-making.

**Limitations**

While the results of this study are promising, several limitations were identified, which should be considered in future research and system development.
- Image quality and ambiguity - poor-quality or ambiguous images submitted by citizens occasionally led to misclassifications or required manual intervention for validation. Enhancements in preprocessing techniques, such as brightness adjustment and noise reduction, can improve model performance in such scenarios. Additionally, public education campaigns on submitting high-quality images could mitigate this issue.
- High-resolution image processing - processing high-resolution images (e.g., 4K) resulted in slightly increased computational demands and longer processing times. Although resizing images during preprocessing alleviated this issue without significant accuracy loss, further optimization is necessary to handle such data more efficiently.
- The current study acknowledges that the reported accuracy (92.8%) is based on a single train-validation split and may not fully capture the model's generalization ability. This limitation is particularly relevant given the constrained variety of conditions in the dataset (e.g., specific urban characteristics, lighting, and weather conditions). A truly independent test set was not used in this study due to the dataset's size and the exploratory nature of the research. Future iterations of this work will prioritize the inclusion of a separate test dataset to provide a more reliable assessment of the model's performance under unseen conditions.
- System scalability - the Dell XPS 9510 laptop performed well for the workloads but exhibited thermal throttling under sustained peak conditions. This limitation suggests the need for enhanced hardware to maintain consistent performance.
- Model generalization - while the model performed well on the provided dataset, its generalization to other municipalities or regions with different urban characteristics remains untested. Expanding training datasets to include diverse conditions and issue types would increase the model's adaptability and robustness.
- Legal and ethical considerations - automated citizen communication must be carefully crafted to avoid misinterpretation or errors, particularly in legal contexts. Further refinements in the natural language processing modules used for citizen feedback are needed to ensure clarity and compliance with administrative standards.
- Hardware dependence - while the Dell XPS 9510 laptop demonstrated sufficient computational capabilities for this study, municipalities with higher data volumes may require more powerful hardware or even distributed computing systems to maintain efficiency.

**Future research**

Firstly, the author emphasizes that this study is intended as a proof-of-concept rather than a fully operational solution for municipalities. Achieving a final-release version would require collaborative efforts from a dedicated team to develop and refine the proposed solution.

Secondly however, to address the above limitations, future research should focus on several key areas such as:
- Model optimization - enhancing the ML model's efficiency and accuracy, particularly for ambiguous or high-resolution images, through advanced algorithms and preprocessing methods.
- Broader testing - deploying the system in diverse municipal contexts to evaluate its adaptability and scalability.
- Citizen education - developing guidelines or educational materials to assist citizens in submitting clear and relevant data.

- Infrastructure improvements - exploring hardware upgrades or cloud-based solutions to handle larger datasets and sustained high workloads.

Despite its limitations, this study establishes a strong foundation for integrating AI-driven image classification into e-government workflows. By addressing the outlined challenges and limitations, future iterations of this system can achieve even greater efficiency, scalability, and citizen satisfaction, paving the way for widespread adoption of advanced e-government technologies.

**Conclusions**

The integration of state-of-the-art ML techniques into e-government workflows exemplifies the transformative impact of AI on public service delivery. This study showcases how AI enhances administrative efficiency and responsiveness by identifying urban issues, generating reports, and providing timely feedback to citizens with high accuracy and performance using widely available hardware.

The results confirm that the proposed model successfully processed images of urban problems, including potholes, waste, illegal parking, and damaged infrastructure, with a classification accuracy of 93%. The system demonstrated the ability to complete the entire workflow, from image analysis to generating actionable reports and citizen notifications, in an average of 7 seconds per case. This time efficiency marks a drastic improvement over traditional manual processes, which typically take up to 8 minutes per case, resulting in a 98.6% reduction in processing time.

The automated system not only streamlined workflows but is capable to enhance citizen engagement by providing immediate and personalized responses. Citizens could receive confirmation of their submissions, details of the actions initiated, and updates on the status of their reports, fostering transparency and trust in government operations. Simultaneously, the system is capable of generating detailed, actionable reports for municipal departments, formatted to administrative standards and ready for implementation.

In addition to its operational efficiency, the system demonstrated scalability, processing up to 500 cases per hour on the Dell XPS 9510 laptop. This performance underscores its potential to handle high volumes of reports, particularly during peak periods such as post-disaster scenarios. The integration of metadata analysis further amplified its utility, enabling the generation of heatmaps and data-driven insights to prioritize resources and inform urban planning strategies.

The adoption of this technology also highlights significant labor optimization benefits. By automating repetitive tasks, the system reduced reliance on large administrative teams, allowing human resources to focus on complex decision-making and strategic initiatives. This shift not only improved the quality of public services but also reduced operational costs, making the solution a viable and cost-effective option for municipalities.

Despite its advantages, the study identified certain challenges. Processing high-resolution or low-quality images occasionally led to delays or misclassifications, which required manual intervention. These limitations emphasize the importance of continuous improvements in preprocessing techniques and citizen education on submitting high-quality images. Furthermore, the system's sustained operation under peak loads occasionally caused minor thermal throttling, highlighting the need for optimized workload management.

Overall, this study illustrates the transformative potential of integrating ML-powered image processing into e-government systems. The demonstrated improvements in efficiency, accuracy, and citizen satisfaction underscore the feasibility of deploying such solutions for enhanced urban governance. By leveraging widely available hardware, municipalities can implement scalable and cost-effective systems that address urban challenges proactively. Future research should focus on optimizing the model for faster processing, expanding its capabilities to address more complex issues, and enhancing its robustness in diverse operational conditions. The adoption of such systems marks a pivotal step toward realizing the vision of e-government 3.0, characterized by transparency, efficiency, and citizen-centric governance.